\documentclass{aa}
\usepackage{graphicx}
\usepackage{txfonts}
\begin{document}
   \title{A near-infrared survey of the entire R Corona Australis
     cloud\thanks{Table A1 is only available in electronic form
       at the CDS via anonymous ftp to cdsarc.u-strasbg.fr (130.79.128.5)
       or via http://cdsweb.u-strasbg.fr/cgi-bin/qcat?J/A+A/}
     }
   \author{
          Martin Haas
          \inst{1}
          \and
          Frank Heymann
          \inst{1}
          \and
          Isabelle Domke
          \inst{1}
          \and
          Holger Drass
          \inst{1}
          \and
          Rolf Chini
          \inst{1}
          \and
          Vera Hoffmeister
          \inst{1}
          }


   \institute{Astronomisches Institut, Ruhr--Universit\"at Bochum,
              Universit\"atsstra{\ss}e 150, 44801 Bochum, Germany
   }

   \date{Received 11. April 2008; accepted 25. June 2008}


  \abstract
{}
{To understand low- to intermediate-mass
star-formation in the nearby R\,Cr\,A molecular cloud, we try to
identify the stellar content that is accessible with near-infrared
observations. }
%
{We obtained a $JHK_{s}$ band mosaic of $\sim 10\arcmin \times
60\arcmin$ covering the entire R\,CrA molecular cloud with
unprecedented sensitivity. We present a catalogue of about 3500
near-infrared sources fainter than the saturation limit $K_{s} \sim
10\,$mag, reaching  $K_{s} \sim$ 18\,mag.
We analysed the extended sources by inspecting their
morphology and point sources by means of colour-colour and
colour-magnitude diagrams. Additionally, we compared the extinction
inferred from the NIR data with the line-of-sight dust emission at
1.2\,mm. Sources towards high dust emission but relatively low
$H-K_{s}$ show a projected mm-exces; these sources are either
immediately surrounded by cold circumstellar material or,
if too red to be a true foreground object, they are embedded in the front 
layer of the 1.2\,mm emitting dust cloud. 
In both cases they are most likely associated with
the cloud.}
%
{By means of the projected mm-excess technique
  we find 33 new faint near-infrared sources
  deeply embedded in the Coronet cluster around R\,CrA,
  for which so far about 20 bright infrared stars have been known.
  In contrast to the Coronet region,
  both the northwestern dust ridge and the southeastern
  cloud condensation "C"
  appear to be devoid of associated stars detectable with
  our near-infrared data.
  Furthermore, about
  a dozen sources, which are spread over the entire molecular cloud region,
  exhibit a possible $K$-band excess, but only
  with marginal statistical significance ($<3\sigma$), so that we do not
  consider the indicated $K$-band excess as real. 
  Finally, while the
  Herbig-Haro-like objects seen
  on our maps are concentrated around the Coronet,
  we find four new nebulae also located farther down to the southeast.
  At the position of IRAS\,18595-3712, an X-shaped bipolar nebula is
  resolved; its exciting star is hidden behind an edge-on disc.
}
{The deep near-infrared
survey of the entire R\,CrA molecular cloud
strengthens the evidence for the Coronet being the
region where most of the young stars are found.
Our results are consistent with earlier predictions
that the R\,CrA cloud has fragmented into
sub-condensations at different star-forming stages.
 }

   \maketitle
%

\section{Introduction}
At a distance of about 170\,pc (Knude \& Hog 1998), the R\,Cr\,A
molecular cloud (Rossano's cloud A, 1978) is among the closest
star-forming regions.
Surveys of this cloud have revealed a population of young
intermediate- to low-mass stars, through the presence of H$\alpha$
emission (e.g. Marraco \& Rydgren 1981, 
Fern\'andez \& Comer\'on 2001) 
or as
X-ray sources (e.g. Forbrich \& Preibisch 2007).
Near-infrared (NIR) observations
detected the {\it Coronet}, a small cluster of embedded
sources around R\,Cr\,A itself (Taylor \& Storey 1984).
Further NIR observations by Wilking et al. (1997)
focussed on the {\it Coronet} region and analysed sources down to
$K = 15\,$mag, which -- due to the large extinction in most parts 
of the observed region -- yields a mass detection limit of about
1\,M$_{\odot}$.
More deeply embedded sources have been identified by radio continuum
(Brown 1987), mid-infrared (Olofsson et al. 1999), and millimetre
(Chini et al. 2003, Nutter et al. 2005) observations.
The young 
(10$^{\rm 3}$ -- 10$^{\rm 5}$~yrs) 
stars drive various
outflows that are observed either as Herbig-Haro objects
(e.g. Hartigan \& Graham 1987, Wang et al. 2004) or as
molecular outflows (e.g. Anderson et al. 1997).
Using the Swedish ESO Submillimetre Telescope (SEST), 
the gas and dust content of the entire R\,Cr\,A cloud having
an extent of 10$\arcmin$\,$\times$\,50$\arcmin$ has been mapped
in C$^{\rm 18}$O J=1-0 by Harju et al. (1993)
and at 1.2\,mm with SIMBA (Chini et al. 2003).
At galactic coordinates $l = 0^\circ, b=-18^\circ$ any fore- and 
background contamination should be negligible.

The R\,Cr\,A molecular cloud lies at the apex of a several
degrees large cometary-shaped flocculent reflection nebula.
As discussed by Harju et al. (1993), the
cloud may have been influenced about 10-13\,Myr ago
by the expanding H\,I shell of the Upper
Centaurus-Lupus OB association located west-northwest of R\,CrA.
Then a rotating disc-like cloud seen roughly edge-on
could have fragmented into several sub-condensations that further
contract to form stars.
The central condensation A, the {\it Coronet},
houses the currently observed young
sources, while the neighbouring regions northwest (NW), and
southeast (C) show virtually no ongoing star formation.
The dense region C was predicted to be a site
for future star formation (Harju et al. 1993), and in fact a
cold cloud core has been detected there (Chini et al. 2003).
On the other hand, it is puzzling why the dust ridge northwest of
the {\it Coronet} lacks signs of recent star formation.
This stimulated our aim of identifying the stellar content along the 
entire cloud down to at least 0.1 M$_{\odot}$ 
(spectral type M7V - M8V, see Landolt-B\"ornstein 1982).

Because of
the expected high extinction at optical wavelengths, we
performed a deep and spatially complete NIR survey of the R\,Cr\,A
cloud.
We gave special emphasis on covering the outermost regions NW and
C, which in the NIR had only been surveyed by 2MASS, and
to map the {\it Coronet} with at least 2 mag higher sensitivity
than was achieved by Wilking et al. (1997).
Here we present the data and perform a first analysis with
respect to the 1.2\,mm dust continuum map. Further comparison of the
NIR data with MIR data will be provided in a follow-up
publication.

\section{Observations and data reduction}

\begin{figure}[htb]
   \centering
   \includegraphics[width=\columnwidth]{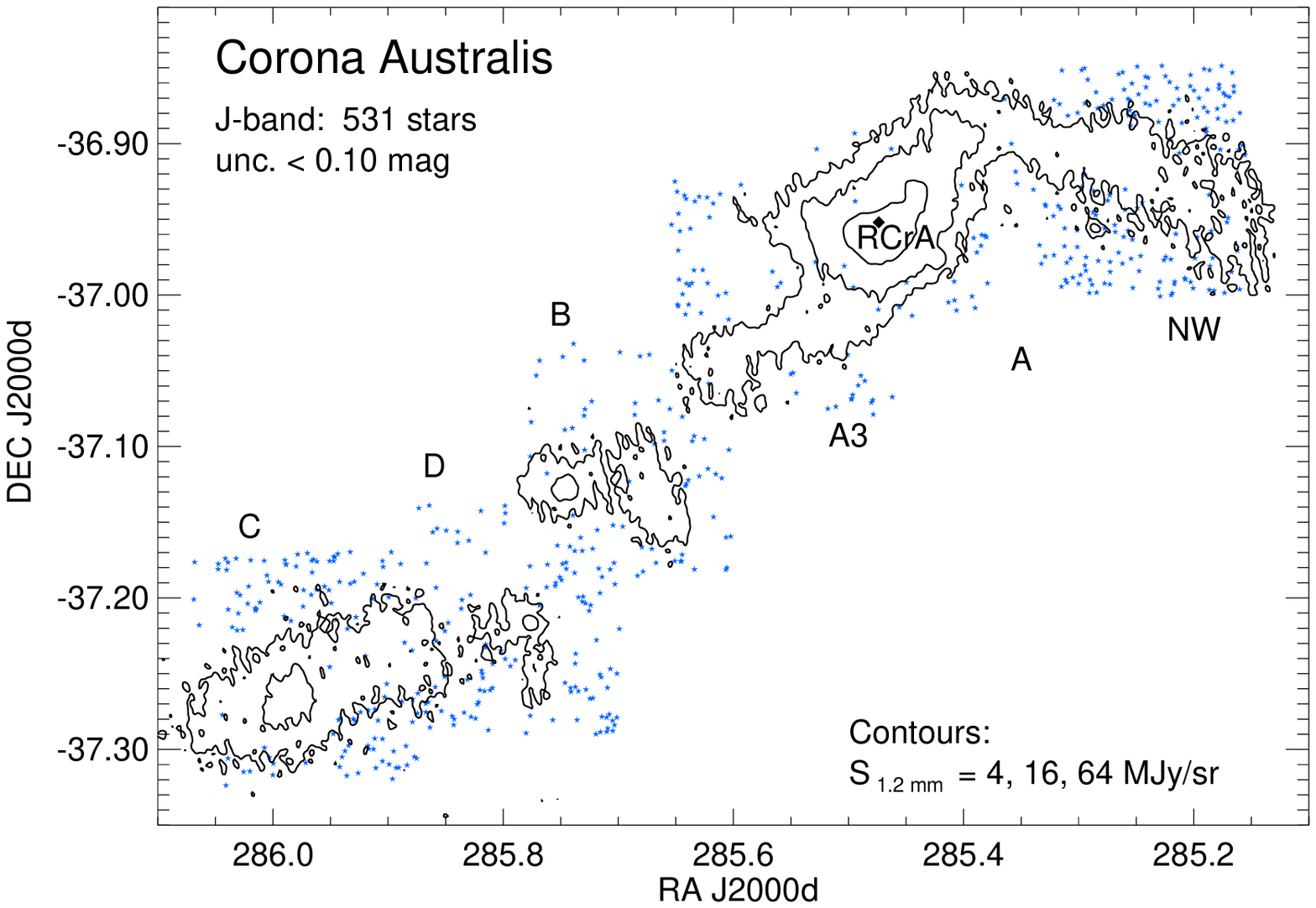}
   \includegraphics[width=\columnwidth]{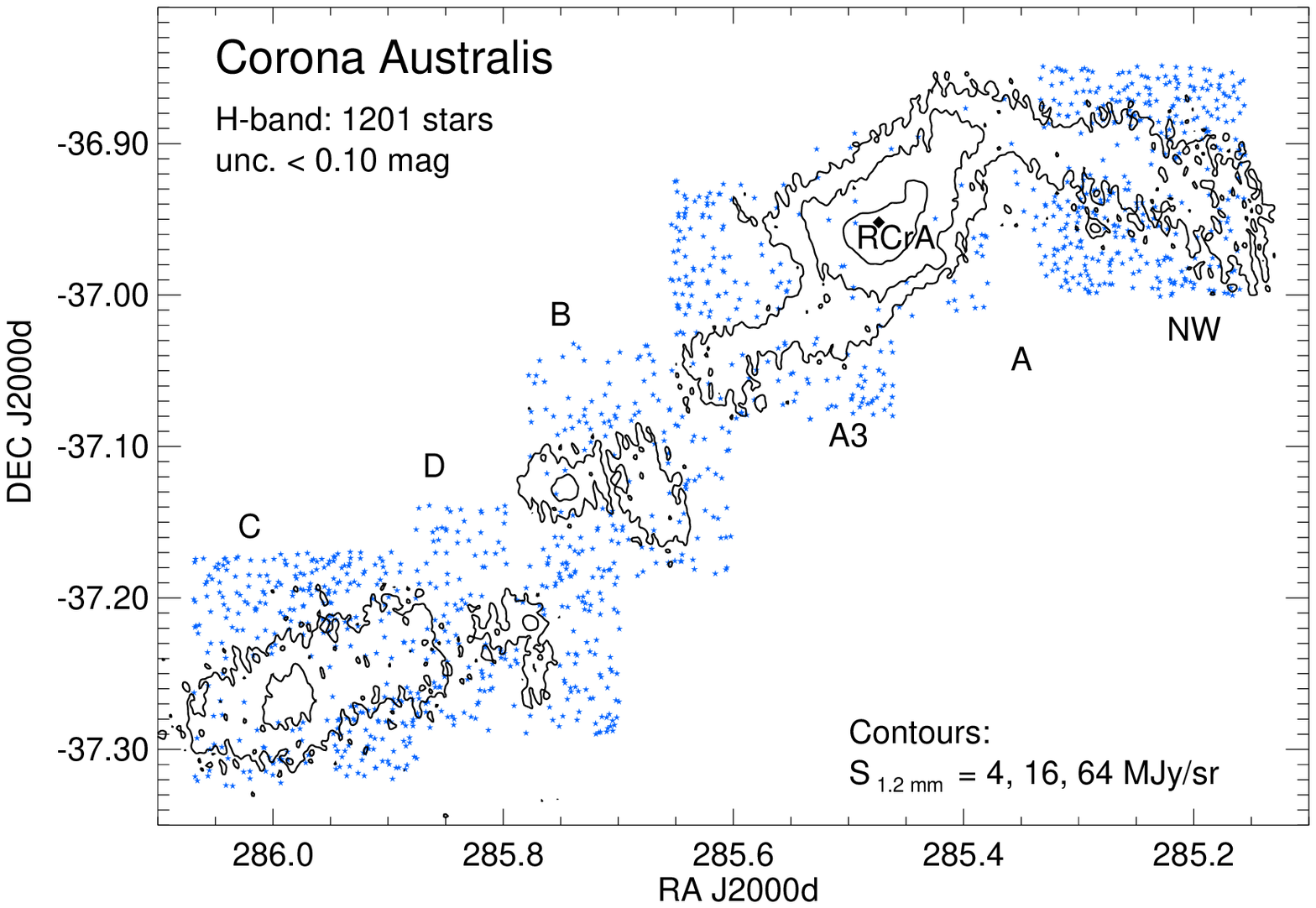}
   \includegraphics[width=\columnwidth]{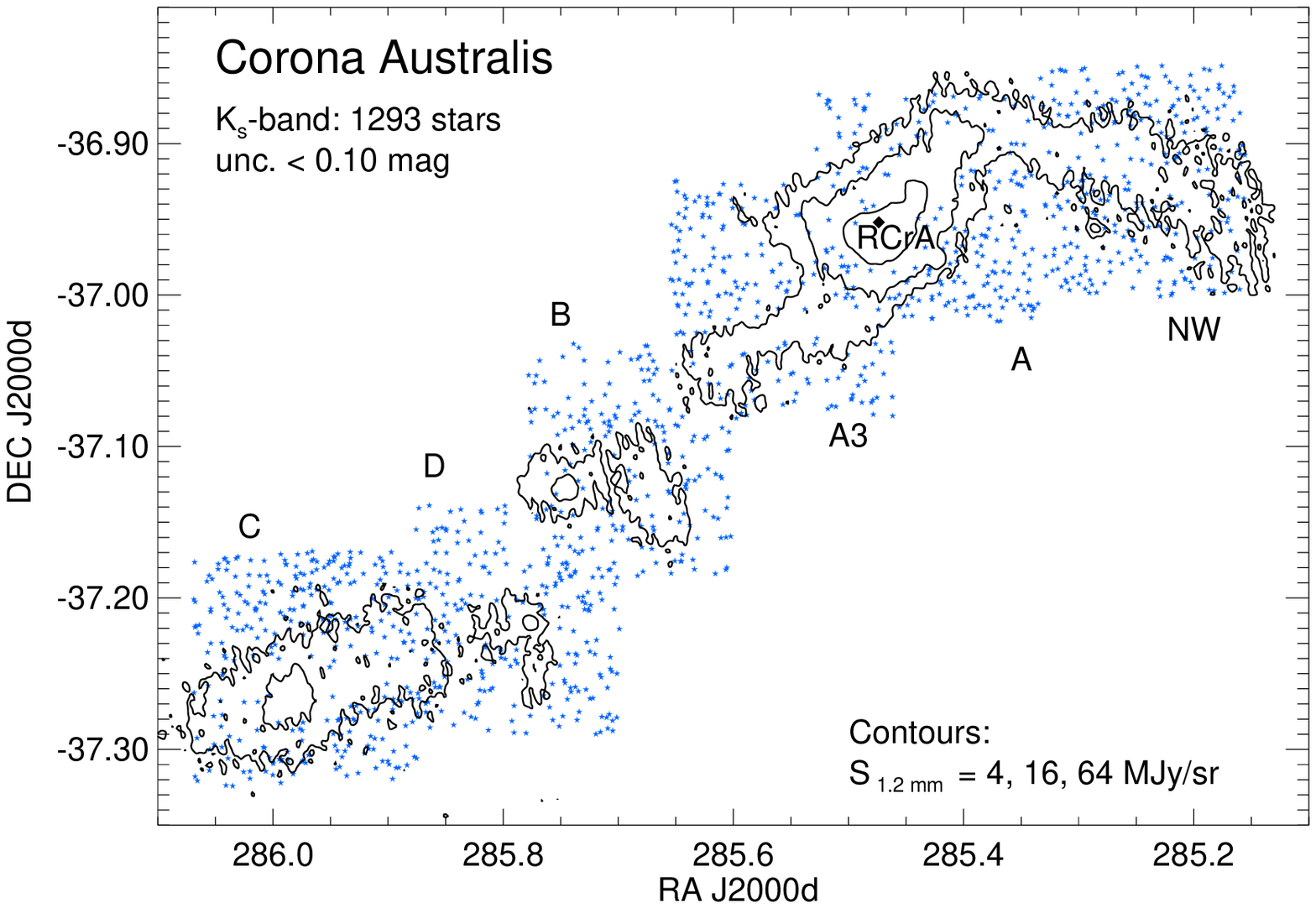}
   \caption{Distribution of IR sources with uncertainty 
     $rms < 0.1\,$mag towards the R\,Cr\,A cloud.
     The 1.2\,mm contours from the SIMBA map are superimposed.
     The fields are labelled (NW, A, A3, B, D, C).
     R\,Cr\,A in the overlap of fields A and A3 is marked with a 
     diamond.
   }
   \label{fig_jhk_mosaics}
\end{figure}

Using ISPI at the CTIO 4-m Blanco telescope with 10$\arcmin$ FOV
 at a pixel size of 0$\farcs$3, we
obtained $J$,$H$,$K_{s}$ 
maps 
of six fields labeled NW, A,
A3, B, D, C covering the entire R\,Cr\,A molecular cloud
(Fig.\,\ref{fig_jhk_mosaics}).

The $J$ and $H$-band observations of field A, requiring long
integrations under good conditions,
were lost due to poor weather, even during a second observing
campaign in 2005. While limited NIR colour information could be derived
for the {\it Coronet} from the overlap with the
neighbouring fields A3 and NW and
from the $J$ and $H$-band data by Wilking et al.(1997), the deep
$K_{s}$-band image of field A allowed us to perform morphological
studies and will be useful for future comparisons, e.g. with the
Spitzer $3.6 - 24\,\mu$m maps 
that cover our fields A, A3, and B, but only 50\% of fields D
and NW
(PI Giovanni Fazio, Lori Allen et al.
in preparation). Essentially each source seen on the Spitzer 3.6 and
4.5\,$\mu$m map was also detected in our $K_{s}$-band image of field
A. 

\begin{table}
\begin{minipage}[t]{\columnwidth}
\caption{Observational parameters (top) and sensitivities (bottom).}
\label{table_obs}
\centering
\renewcommand{\footnoterule}{}  
\begin{tabular}{crrcc}
Field & RA J2000d & Dec J2000d & Field size  & Obs. date  \\
\hline\hline
  C & 285.971985 & --37.246700 & 9$\farcm$5 $\times$ 9$\farcm$3 & 31 July 2004 \\
  D & 285.788513 & --37.214447 & 8$\farcm$6 $\times$ 9$\farcm$1 & 29 July 2004 \\
  B & 285.689514 & --37.108849 & 8$\farcm$5 $\times$ 9$\farcm$2 & 29 July 2004 \\
 A3 & 285.559021 & --37.003448 & 9$\farcm$4 $\times$ 9$\farcm$5 & 29 July 2004 \\
  A & 285.429504 & --36.941799 & 9$\farcm$4 $\times$ 9$\farcm$1 & 30 July 2004 \\
 NW & 285.247498 & --36.925301 & 8$\farcm$9 $\times$ 9$\farcm$2 & 29 July 2004 \\
\hline
  &  &  &  \\
\end{tabular}
\begin{tabular}{@{\hspace{1mm}}c@{\hspace{1mm}}ccccc@{\hspace{1mm}}}
\hline\hline
Field & Filter &  Integr. & Detect. & rms$<$0.2 &Complete.\\
      &        &   time [s]           & limit [mag]      & limit [mag]    & limit [mag] \\
\hline
C  & $J$    & 2000 &20.3 & 19.0 &  17.6 \\
C  & $H$    &  270 &19.1 & 18.3 &  17.1 \\
C  & $K_{s}$&  120 &18.4 & 17.6 &  16.8 \\
D  & $J$    &  375 &19.1 & 18.4 &  17.1 \\
D  & $H$    &  270 &18.8 & 18.1 &  16.8 \\
D  & $K_{s}$&  120 &17.7 & 17.1 &  16.3 \\
B  & $J$    &  375 &18.8 & 18.3 &  16.9 \\
B  & $H$    &  270 &18.2 & 17.8 &  17.0 \\
B  & $K_{s}$&  120 &17.5 & 17.2 &  16.4 \\
A3 & $J$    &  375 &19.3 & 18.7 &  18.0 \\
A3 & $H$    &  270 &19.0 & 18.5 &  17.1 \\
A3 & $K_{s}$&  120 &17.6 & 17.3 &  16.2 \\
A  & $K_{s}$& 6800 &18.2 & 17.7 &  16.9 \\
NW & $J$    & 1000 &19.4 & 18.5 &  17.3 \\
NW & $H$    &  600 &19.0 & 18.3 &  17.0 \\
NW & $K_{s}$&  120 &17.2 & 16.8 &  16.0 \\
\hline
\end{tabular}
\end{minipage}
\end{table}

\begin{table*}[b!]
\begin{minipage}[t]{\textwidth}
\caption{Extended objects. Positions and photometry from suitably
  chosen apertures, errors are
given in brackets.} \label{table_nebulae} \centering
\begin{tabular}{rrcccccc}
\hline\hline
RA J2000d & DEC J2000d & $J$ [mag] & $H$ [mag] & $K_{s}$ [mag] & S1.2\,mm [MJy/sr] & Remarks, size ["]                  \\
\hline
285.508212  &  -36.899951  &  ...            &  ...          & 15.80 ~(0.18) & -0.8  ~(1.1) & Star with nebula$^{a}$, $\sim$3$\arcsec$ \\
285.512850  &  -37.002804  &  ...            &  16.32~(0.09) & 14.34 ~(0.15) & 11.8  ~(2.2) & Star with nebula$^{a}$, $\sim$3$\arcsec$  \\
285.522940  &  -36.910680  &  ...            &  ...          & 11.60 ~(0.30) & -0.3  ~(0.9) & HH\,99$^{a}$ , $\sim$25$\arcsec$           \\
285.552670  &  -36.945956  &  16.26 ~(0.09)  &  15.20~(0.08) & 13.78 ~(0.08) &  7.7  ~(1.8) & HH\,735, $\sim$10$\arcsec$     \\
285.612310  &  -37.175960  &  16.64 ~(0.06)  &  15.09~(0.05) & 14.38 ~(0.05) &  0.1  ~(0.7) & Star with nebula, $\sim$5$\arcsec$  \\
285.702890  &  -37.238030  &  16.82 ~(0.09)  &  15.24~(0.09) & 14.74 ~(0.04) &  0.9  ~(1.7) & Star with nebula, $\sim$4$\arcsec$  \\
285.744250  &  -37.126140  &  ...            &  14.32~(0.10) & 12.60 ~(0.07) &  32.8 ~(7.3) & Isabelle Nebula, MMS\,23, $\sim$20$\arcsec$   \\
285.883453  &  -37.286964  &  15.32 ~(0.15)  &  14.12~(0.15) & 13.91 ~(0.15) &  0.4  ~(0.6) & Star with nebula, $\sim$5$\arcsec$   \\
\hline
\end{tabular}\\
\end{minipage}
$^{a}$ also seen by Wilking et al. (1997)
\end{table*}

The ISPI integration times per field were adjusted to the
extinction estimated from the 1.2\,mm SIMBA map (Chini et al. 2003).
The observational parameters and the resulting sensitivities are
listed for each field in Table \ref{table_obs}. 
The field size refers to the area covered by all three filters. 
Because the observations at different filters had
small offsets, the resulting common field sizes differ slightly from
field to field. 
The detection limit, as well as the limiting sensitivity of sources with
rms$<$0.2 mag, refer to the mean of the faintest three objects in the
field-filter combination.
We determined the 100\% completeness limit for each field and filter 
from the magnitude histograms. The completeness
limits are uncertain, because the maps 
are not homogeneous; for example near R\,CrA large areas are saturated or
affected by spikes.

Each mosaic has been observed with several dither cycles
(20$\arcsec$) under moderately photometric but
good seeing conditions (1$\arcsec$ -- 1$\farcs$5).
We did not collect extra sky frames, but instead constructed the
mean sky frames from the field images themselves. 
Therefore, our photometry is limited to objects with an extent $<
20\arcsec$. Dome flats were taken at the beginning and end of each
night.

Standard data reduction was performed using the software packages
for the ISPI instrument implemented in 
the Image Reduction and Analysis Facility 
(IRAF). This includes 
astrometric corrections across each field. The sources were
extracted using the DAOFIND and DAOPHOT tools and for comparison
using the SEXTRACTOR tool (Bertin \& Arnouts 1996). The astrometric and
photometric calibration was performed using 2MASS stars in the
fields. The astrometric accuracy was 
conservatively estimated to be
better than 1$\arcsec$, while
the photometric uncertainty depends on the brightness of the
sources; we applied a cut of $rms = 0.4\,$mag for the final
catalogues. For sources located in the overlap of two adjacent fields
we list the mean photometry from both fields.

We inspected the individual frames for sources with extended
morphology. They are listed in Table\,\ref{table_nebulae},
with photometry derived
from suitably chosen apertures. Point sources are listed with PSF
photometry in Table A1, available only in electronic form, 
and having the same layout as Table\,\ref{table_nebulae}.
Sources
brighter than $K_{s} \sim 10\,$mag are saturated (outside the linear
range of the detector) and therefore
excluded. Tables\,\ref{table_nebulae} and A1
also list the SIMBA surface brightness
$S_{1.2\,mm}$ averaged over $28\arcsec \times 28\arcsec$ at the sky
position of the sources (SIMBA HPBW = 24$\arcsec$). 
The SIMBA surface brightness does not necessarily 
refer to isolated 1.2\,mm point sources, rather it is derived from the
mostly broad and smooth 1.2\,mm emission towards each NIR source, in
order to provide a measure of the dust column density in the
direction of the star. The dust producing this emission may be
located in front, nearby, or behind the star.   

\begin{figure}
   \centering
   \includegraphics[width=\columnwidth]{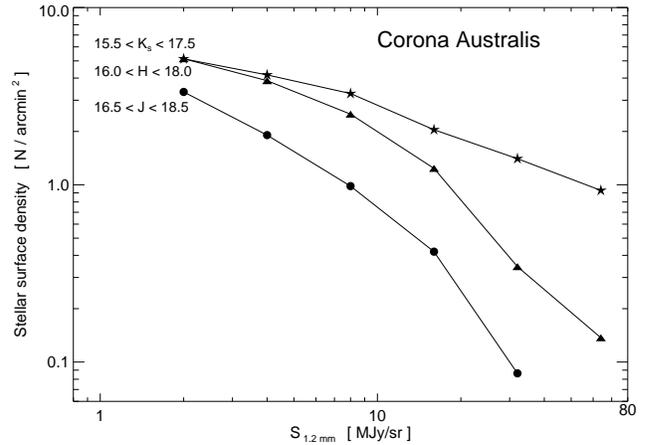}
   \caption{Surface density of sources with $rms < 0.2\,$mag
     as a function of wavelength and the SIMBA 1.2\,mm emission
     towards these sources. For $J$ and $H$, only five fields are
     considered, while $K_{s}$ refers to all six fields including
     field A.
   }
   \label{fig_jhk_surface_density}
\end{figure}

\section{Results and discussion}

Figure \ref{fig_jhk_mosaics} shows for each filter $J$, $H$, and $K_{s}$
the sky position of the point sources having photometric $rms <
0.1\,$mag, as well as the SIMBA 1.2\,mm contours superimposed. The most
striking result from Fig.~\ref{fig_jhk_mosaics} is the steep decline
in source density towards sky regions with bright 1.2\,mm emission,
i.e. highest dust column density. This is illustrated in
Fig.~\ref{fig_jhk_surface_density}. The decline in stellar surface
density is most prominent in the $J$-band and steepens considerably
in the $H$-band above $S_{1.2} >  20$\,MJy/sr; the $K_{s}$-band
appears less affected. However, even at 2.2\,$\mu$m the dusty
central regions of the R\,Cr\,A cloud, in particular towards fields
A and C, turn out to be opaque for background stars, dimming them below
$K \sim 18\,$mag. Stars seen towards these opaque regions may be
foreground stars or located within the front layers of the cloud. On
the other hand, most of the sources towards low 1.2\,mm surface
brightness are likely background stars. 

To identify stars associated with the R\,Cr\,A cloud, we
applied three strategies:

\begin{itemize}
\item Projected millimetre excess sources
  (Sect. \ref{section_extinction_deficient_sources}),
\item $K_{s}$-band excess sources (Sect. \ref{section_disc_excess_sources}),
\item Nebulae and extended objects (Sect. \ref{section_nebulae}).
\end{itemize}

\subsection{Projected millimetre excess sources}
\label{section_extinction_deficient_sources}

To determine the cloud membership for those point sources
with $H$ and $K_{s}$-band photometry available, and independent of
$J$-band detection, we apply a new strategy here involving the
1.2\,mm data. Basically we follow the well-known fact that, for a
background source, the extinction derived from the dust column, i.e.
the 1.2\,mm brightness, should match the $H-K_{s}$ colour of that
source.
The observed reddening (extinction)
is correlated with the hydrogen column density, and -- 
via the dust/gas ratio -- also with the dust column density 
($A_{V} / N_{H} \sim 5.6\,\times\,10^{-22} mag/cm^{-2}$ --- 
Seward 1999). 
We trawl the cloud members using the
cloud's dust emission as a fishing net: if a source lies towards a
high dust column but has too a small "deficient" extinction as
derived from $H-K_{s}$, then this source cannot be located behind
the dust cloud, but instead it must be a foreground source or a cloud
member. 
Such a "projected mm-excess" source is either
surrounded by cold circumstellar material typical for a very young 
object likely associated with the parent molecular cloud,
or it must be embedded in the front layer of the
dust cloud, 
if the $H-K_{s}$ derived extinction is too high for a
true foreground source. 
The projected mm-excess technique is particularly 
suited to identifying cloud sources without $K$-band excess.

\begin{figure}
   \centering
   \includegraphics[width=\columnwidth]{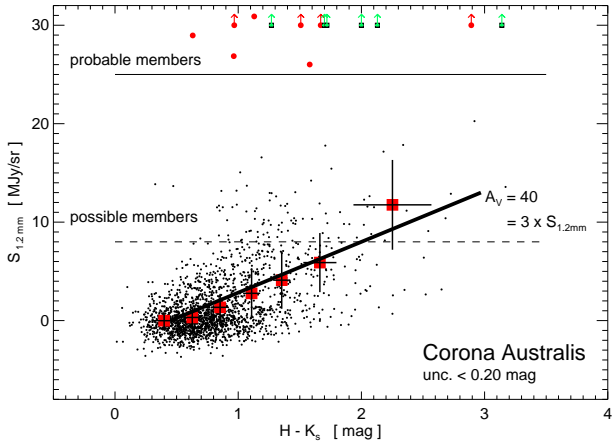}
   \caption{1.2\,mm surface brightness versus $H-K_{s}$.
    Sources above the solid line $S_{1.2} = 25$\,MJy/sr are
    considered to be probable members, and sources above the long-dashed
    line $S_{1.2} = 8$\,MJy/sr are called possible members.
    The thick solid line indicates $A_{V}$ adopting standard dust
    properties; it is not a fit to the median data points (red
    squares).
   }
   \label{fig_extinction_deficient}
\end{figure}

Figure \ref{fig_extinction_deficient} shows the distribution
of the 1.2\,mm surface brightness versus $H-K_{s}$. Here we plot all
sources with photometric uncertainty $<$ 0.2\,mag.
The bulk of the sources lie at both low $S_{1.2} < 5$\,MJy/sr
and $H-K_{s} < 1$, consistent with moderately reddened
background stars off the cloud cores.
There are also many objects exhibiting larger $H-K_{s}$
and considerable $S_{1.2} > 8$\,MJy/sr. These objects are seen in
projection towards the dense dust cloud.
The red squares show the median of $S_{1.2}$ along $H-K_{s}$ bins (with
upper and lower quartiles as error bars). The thick solid line
represents the extinction $A_V$ predicted from the dust column,
using standard dust properties ($\kappa$~=~0.37\,cm$^{\rm 2}$g$^{\rm
-1}$, Kr\"ugel 2003). The results derived below are quite
independent of the detailed dust properties. The $A_V$ line is not a
fit to the median data points, but agrees well with them up to
$H-K_{s} < 1.5$. At higher $H-K_{s}$, the agreement of the data with
the $A_V$ line improves after omitting the stars with extremely high
$S_{1.2} > 25$\,MJy/sr.

In the following we distinguish between sources with extremely bright
$S_{1.2} > 25$\,MJy/sr, which we call {\it probable members}, and sources
with intermediate $S_{1.2} >$ 8\,MJy/sr, which we call {\it 
possible members}. 

\subsubsection{Probable members}
\label{section_cloud_members}
There are only two spots outside of the {\it Coronet} that have
$S_{1.2} > 25$\,MJy/sr: VV\,CrA (MMS\,24) and MMS\,23
(see Sect.~\ref{section_nebulae}) located in fields D and B,
respectively.
The dust emission in these bright spots ($S_{1.2} \sim 30$\,MJy/sr)
is probably directly attributed to the PMS stars
(Wilking et al. 1992; Chini et al. 2003; Nutter et al. 2005).
The source IRS\,2, coincident with MMS\,9, is
saturated ($K=7.16$ mag).
These three sources are not considered further here.

\begin{figure}
   \centering
   \includegraphics[width=\columnwidth]{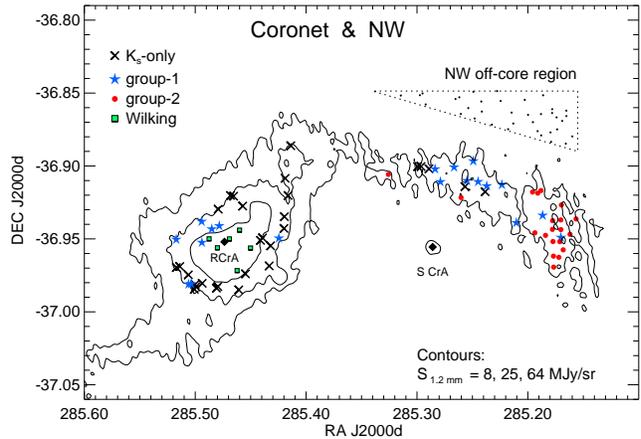}
   \caption{Sky position of the probable members and the possible NW cloud
     members.
     All probable members are located
     towards the {\it Coronet} around R\,CrA (filled diamond).
     Some sources lie near saturated areas or spikes
     in our $K_s$-image, and their
     photometry was taken from Wilking et al. (1997).
     The possible members are located towards the core of the
     NW dust ridge, north and west of S\,CrA (filled diamond).
     The subdivision of stars into group-1/-2 has been derived from
     the colour-magnitude diagram (Fig.\,\ref{fig_field_nw}).
     The dotted triangle encircles the stars in the NW off-core region
     used for comparions.
   }
   \label{fig_coronet_nw}
\end{figure}

There are 14 sources with $S_{1.2} > 25$\,MJy/sr far
above the $A_V$ line in Fig.~\ref{fig_extinction_deficient}.
All of them are located towards the {\it Coronet}
(Fig.~\ref{fig_coronet_nw}). In our $K_{s}$ image, the immediate
neighbourhood of R\,CrA is crowded with bright nebulous features that 
make any detection of faint sources difficult. Therefore, our new
detected $K_{s}$ sources only appear at some distance from R\,CrA.
The 14 probable members are not distinct 1.2\,mm point sources in
the SIMBA map, but their 1.2\,mm flux is due to extended cloud
emission towards their line of sight.
If they were background stars behind the bright 1.2\,mm dust emission,
their light would suffer an extinction of $75 < A_V < 500$.
In contrast, the extinction derived from $H-K_{s}$ is much
lower ($2 <  A_V < 40$), but too high for a true foreground source.
Therefore, they must be embedded in the front layer of the cloud.

We have also considered the location of 
known members of the R\,CrA association in 
Fig.~\ref{fig_extinction_deficient}.
Their membership has been inferred, for instance, 
from  H$\alpha$ (Marraco \& Rydgren 1981; Fernand\'ez \& Comer\'on 2001) 
or X-ray emission (Forbrich \& Preibisch 2007). 
These sources are not exclusively
located towards regions with high 1.2\,mm dust emission, 
but many of them lie in regions around the cloud having a 
low projected dust column. 
Also, the known members exhibit a wide range of extinction 
properties (range in $H-K_{s}$). From this it is clear that 
the known members are more or less randomly distributed thoughout 
the entire range depicted in Fig.~\ref{fig_extinction_deficient}. 
We did not overplot them in this figure with extra symbols,  
to avoid confusion with the other symbols. While known members 
are located both towards and around the cloud, 
the projected mm-excess technique is limited to identifying only the
sources located in projection against the cloud.    

\begin{figure}
   \centering
   \includegraphics[width=\columnwidth]{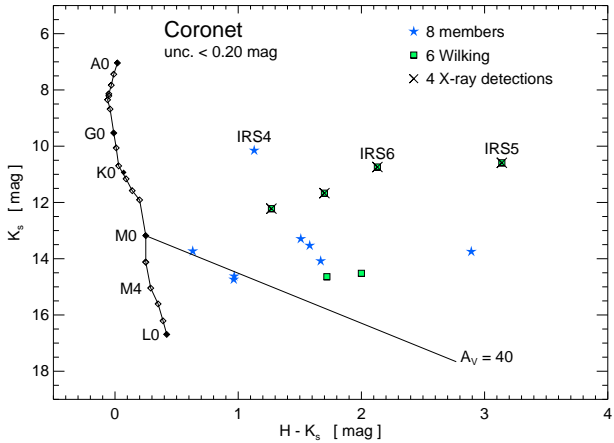}
   \caption{Colour-magnitude diagram of the probable members,
   i.e. projected mm-excess sources with $S_{1.2} > 25$\,MJy/sr.
   }
   \label{fig_col_mag_members}
\end{figure}

Figure~\ref{fig_col_mag_members}  shows the $K_{s}$ vs.
$H-K_{s}$ colour-magnitude diagram of the probable members.
For comparison, the unreddened main sequence at 170\,pc
distance and the extinction vector is
plotted\footnote{The main sequence
  was taken from Schmidt-Kaler (1982)
  for spectral type A0 to M4 and from Dahn et al. (2002) for M0 to L0.
  The two references exhibit a discrepancy of about 1\,mag for M0
  stars. To bridge this gap and to provide a smooth
  presentation in the plot, we stretched the Schmidt-Kaler main
  sequence evenly between G0 and M4, fitting the more actual M0 magnitudes of
  Dahn et al. In colour-colour plots, the location of the main
  sequence is virtually not affected.}.
The probable members are
stars of intermediate to low mass down to about spectral type M0
(0.5 $M_{\odot}$, see Landolt-B\"ornstein 1982).
If they had a strong K-band excess, some sources could even have 
lower mass, but we did not find evidence of a strong $K$-excess as shown
in Sect.\,\ref{section_disc_excess_sources}.
Comparing our data with the Chandra results by Forbrich \& Preibisch
(2007), only the brightest stars ($K_{s} < 12\,$mag, all being
young) are detected in the X-ray map. This suggests that deeper
X-ray observations are required to draw further conclusions
about the accretion activity of our faint probable members.
The 9 sources fainter than $K_{s} = 13$ mag were not known to
be associated with the cloud. 
They are not detected by 2MASS.
The mm-excess strategy enables us to establish their cloud
membership. 
The probable members are listed in Table~\ref{table_members}.

\begin{figure}
   \centering
   \includegraphics[width=\columnwidth]{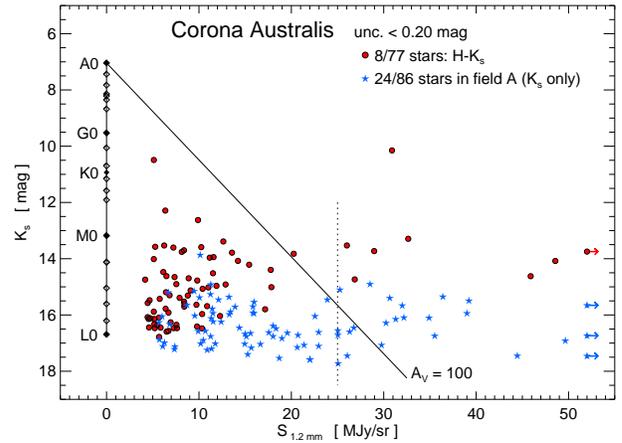}
   \caption{ $K_{s}$ brightness versus $S_{1.2}$:
     Plotted are sources of field A having only  $K_{s}$
     photometry. They are
     compared with sources detected in both $H$ and $K_{s}$ in any of
     our fields,
     which are located towards $S_{1.2}> 25~MJy/sr$ or have
     red $H-K_{s} > 1.5$.
   }
   \label{fig_k_only}
\end{figure}

Next we consider the numerous faint $K_{s}$-only sources 
detected in field A without appropriate $H$-band photometry available.
Figure~\ref{fig_k_only} shows the $K_{s}$
brightness vs. $S_{1.2}$. For comparison the extinction vector and
sources with $H$ detection in the other fields (C, D, B, A3, NW)
are plotted.
There are 24 $K_{s}$-only sources
with projected  mm-excess $S_{1.2} > 25$\,MJy/sr.
Because they have $K_{s}\sim$ 15-18\,mag, our sources
are not seen in deep optical images
($R \sim$ 22.5\,mag, $I \sim$ 20.5\,mag, L\'opez Mart\'i et al. 2005).
The colours $R - K_{s} > 4.5$ and $I - K_{s} > 2.5$
rule out the possibility of unreddened
foreground stars.
If they were located behind the very bright
1.2\,mm dust emission ($S_{1.2}> 25~MJy/sr$ corresponding to $A_{V}
> 75$), they would form an ensemble
of background stars earlier than spectral type A0 with a projected
concentration towards the {\it Coronet}.
Because this is unlikely and not consistent
with our analysis of various off-cloud regions, we conclude that
the 24 newly-found infrared sources with projected  mm-excess
are deeply embedded in the {\it Coronet}.

Compared to probable members with $H$ photometry available,
the $K_{s}$-only sources are fainter (Fig.~\ref{fig_k_only}, 
Table~\ref{table_members})
and probably more reddened.
For an adopted extinction $A_{V} \sim 30-40^{m}$
(see Fig.~\ref{fig_col_mag_members}),
their dereddened brightness would be about $K_{s}$ = 12-15 mag.
If not Herbig-Haro knots, then as (presumably PMS)
K-stars and M-dwarfs, they would represent the lower
mass extension of the known {\it Coronet} members.
Because theoretical concepts and numerical simulations find
that lower mass stars evolve more slowly than higher mass stars,
one may expect that the low-mass end of the $K_{s}$-only sources
will show a strong disc excess at mid-IR wavelengths.

\begin{table*}[ht]
\begin{minipage}[t]{\textwidth}
\caption{Probable members found via the mm excess technique.
Positions and photometry with errors 
given in brackets.} \label{table_members} \centering
\begin{tabular}{rrccccc}
\hline\hline
RA J2000d & DEC J2000d & $J$ [mag] & $H$ [mag] & $K_{s}$ [mag] & S1.2\,mm [MJy/sr] & additional photometry reference               \\
\hline
285.413788 & -36.886112 &     ...      &    ...      &15.11 (0.04) & 25.3 ( 3.3) &         \\
285.416229 & -36.920238 &     ...      &    ...      &17.73 (0.19) & 25.1 ( 1.7) &         \\
285.418884 & -36.908459 &     ...      &    ...      &17.45 (0.19) & 26.1 ( 2.2) &         \\
285.419556 & -36.934746 &     ...      &    ...      &16.74 (0.11) & 35.5 ( 7.2) &         \\
285.419922 & -36.942795 &     ...      &    ...      &16.10 (0.08) & 34.9 ( 4.6) &         \\
285.424561 & -36.949516 & 17.67 (0.14) &14.80 (0.04) &13.29 (0.01) & 32.6 ( 6.2) &        \\
285.432007 & -36.954708 &     ...      &    ...      &15.35 (0.05) & 36.4 ( 5.6) &         \\
285.433258 & -36.968506 &     ...      &    ...      &15.67 (0.05) & 32.0 ( 4.6) &         \\
285.439301 & -36.949272 &     ...      &    ...      &15.66 (0.07) & 63.8 ( 9.8) &         \\
285.441071 & -36.951004 &     ...      &    ...      &17.46 (0.19) & 67.3 ( 9.5) &         \\
285.450104 & -36.956223 & 17.54 (0.30) &13.73 (0.02) &10.59 (0.00) & 92.2 (10.9) & Wilking$^{a}$, IRS5 Taylor \& Storey$^{b}$\\
285.454681 & -36.973965 &     ...      &    ...      &16.74 (0.11) & 60.6 (15.1) &         \\
285.457184 & -36.927448 &     ...      &    ...      &16.92 (0.13) & 49.7 ( 5.7) &         \\
285.460022 & -36.943954 & 16.35 (0.30) &12.87 (0.02) &10.74 (0.00) & 74.6 ( 5.2) & Wilking$^{a}$, IRS6 Taylor \& Storey$^{b}$ \\
285.460724 & -36.984917 &     ...      &    ...      &15.40 (0.07) & 30.7 ( 5.8) &         \\
285.462250 & -36.971722 &     ...      &16.52 (0.18) &14.52 (0.12) &103.2 (16.1) & Wilking$^{a}$ \\
285.465393 & -36.920685 &     ...      &    ...      &14.91 (0.04) & 28.5 ( 5.8) &         \\
285.468170 & -36.920223 &     ...      &    ...      &16.55 (0.09) & 25.0 ( 4.6) &         \\
285.469177 & -36.950024 & 17.77 (0.30) &13.37 (0.02) &11.67 (0.02) & 84.7 ( 9.1) & Wilking$^{a}$, IRS9 Taylor \& Storey$^{b}$ \\
285.478333 & -36.941036 &     ...      &15.59 (0.14) &14.62 (0.03) & 45.9 ( 6.9) &        \\
285.479218 & -36.929649 &     ...      &    ...      &16.72 (0.11) & 25.0 ( 4.4) &         \\
285.479614 & -36.982685 &     ...      &    ...      &17.45 (0.19) & 44.5 (10.0) &         \\
285.480042 & -36.956081 &     ...      &13.49 (0.22) &12.22 (0.14) &170.7 (23.6) & Wilking$^{a}$, IRS7 Taylor \& Storey$^{b}$ \\
285.480865 & -36.983723 &     ...      &    ...      &15.50 (0.05) & 39.2 ( 8.6) &         \\
285.485077 & -36.943222 &     ...      &15.75 (0.18) &14.08 (0.03) & 48.6 ( 9.0) &        \\
285.487518 & -36.949936 &     ...      &16.36 (0.28) &14.64 (0.08) & 87.9 (19.2) & Wilking$^{a}$ \\
285.493683 & -36.980431 &     ...      &    ...      &15.94 (0.07) & 39.0 ( 6.2) &         \\
285.493988 & -36.952465 &     ...      &16.64 (0.08) &13.75 (0.03) &101.8 (18.4) &        \\
285.494202 & -36.938053 & 13.82 (0.30) &11.28 (0.20) &10.15 (0.20) & 30.9 ( 5.4) &        \\
285.499359 & -36.982796 &     ...      &    ...      &17.07 (0.15) & 29.8 ( 5.0) &         \\
285.500153 & -36.981281 &     ...      &    ...      &16.09 (0.08) & 32.2 ( 5.1) &         \\
285.500824 & -36.984539 &     ...      &    ...      &16.46 (0.09) & 26.8 ( 5.1) &         \\
285.503662 & -36.980827 & 16.06 (0.04) &14.36 (0.02) &13.73 (0.03) & 29.0 ( 5.9) &        \\
285.506012 & -36.981091 & 17.74 (0.18) &15.70 (0.06) &14.74 (0.04) & 26.9 ( 5.3) &        \\
285.506165 & -36.974621 &     ...      &    ...      &16.15 (0.08) & 31.3 ( 5.1) &         \\
285.513916 & -36.968899 &     ...      &    ...      &16.28 (0.09) & 30.2 ( 4.2) &         \\
285.517029 & -36.969604 &     ...      &    ...      &16.60 (0.12) & 26.3 ( 3.6) &         \\
285.517120 & -36.950371 &     ...      &15.11 (0.04) &13.53 (0.01) & 26.0 ( 2.2) &        \\
\hline
\end{tabular}\\
\end{minipage}
$^{a}$ Wilking et al. (1997), $^{b}$ Taylor \& Storey 1984
\end{table*}

\subsubsection{Possible members}
\label{section_cloud_candidates}

In contrast to the {\it Coronet} region,
both the northwestern dust ridge and the southeastern
cloud condensation C are much fainter at 1.2\,mm, so that we cannot
simply trawl the associated stars by means of a projected  mm-excess.
As a trial we investigate how far we can extend the technique to
fainter dust surface brightness.
We select possible members having $S_{1.2}>$8\,MJy/sr
(Fig.\,\ref{fig_extinction_deficient}).
One possible member, the bright T\,Tauri star
S\,CrA (MMS\,1) having $S_{1.2} \sim 19$\,MJy/sr,
clearly is a cloud member.
However, we have excluded this star from the following investigation,
because its mm emission comes from a dust
disc around the star and not from a smooth background (Wilking et al
1992, Chini et al. 2003, Nutter et al. 2005).
In addition to the cores of NW and C,
we have analysed several off-core regions in fields C, D, B and NW,
in order to estimate the
number and brightness of virtually unreddened background stars.
All studied off-core regions yield similar number counts
(Table\,\ref{table_stellar_surface_density})
supporting that these regions are dominated by background stars.
In the following we restrict the presentation of 
our analysis to the NW region,
the results for region C are similar.

\begin{figure}
   \centering
   \includegraphics[width=\columnwidth]{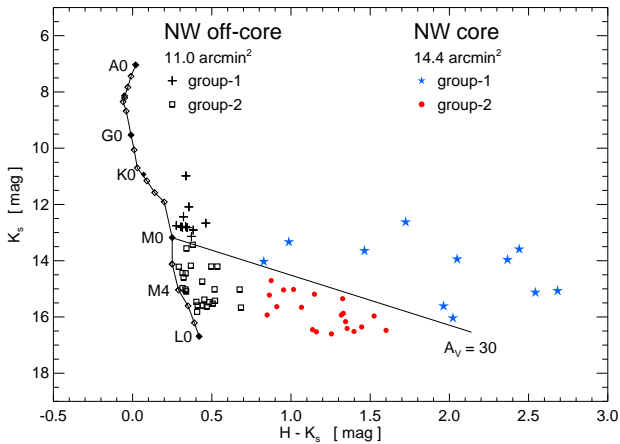}
   \caption{Colour-magnitude diagram of the NW-core and NW-off-core.
     Group-1/-2 sources are defined to lie above/below the $A_V$ line.
     The main sequence shifted to the distance of R\,CrA
     is plotted only for illustrative
     purposes.
   }
   \label{fig_field_nw}
\end{figure}

The possible members lie towards the core of the NW dust-ridge
north and west of S\,CrA
(Fig.~\ref{fig_coronet_nw}).
Figure~\ref{fig_field_nw} shows the colour-magnitude diagram of the
possible members in the NW core and the stars in the NW off-core region.
All possible members have
high enough extinction values ($A_V >$~6\,mag)
derived from $H-K_{s}$ to rule out their being unreddened
foreground stars.
If the possible members were at the distance of R\,CrA, then
they would be low-mass and even very low-mass stars (between spectral
type K0 and M0, smaller than M0, respectively).
However, we have to check that they are not background stars.

For the purpose of this check, we subdivide the NW core
sources into group-1 and group-2 lying above/below the solid
line in Fig.\,\ref{fig_field_nw}
and compare their extinction and sky position.
Group-2 has, on average, smaller $H-K_{s}$, i.e. less
extinction than group-1.
Also, group-1 and group-2 sources populate mainly
the right
and left halves in Fig.\,\ref{fig_extinction_deficient}. 
Thus at a given 1.2\,mm flux as measured by the
SIMBA beam, group-2 sources suffer from less extinction.
Group-1 and $K_{s}$-only sources concentrate
towards the denser eastern region, but
group-2 sources strongly populate the western border of the
core at $RA<285.20$ (Fig.~\ref{fig_coronet_nw}).
This western border exhibits an abrupt change in the dust-to-gas
properties: While the 1.2\,mm emission roughly remains constant,
the C$^{18}$O emission declines rapidly at $RA=285.20$.
A natural explanation is that, at the cloud border,
the dust distribution is clumpy
with low filling factor not resolved in the
SIMBA beam.
Then a background object can easily be seen
as group-2 source through a hole between dust clumps.
This picture is consistent with the expectation that the wind from the
Upper Centaurus-Lupus OB association corrodes the cloud's western
front.

\begin{figure}
   \centering
   \includegraphics[width=\columnwidth]{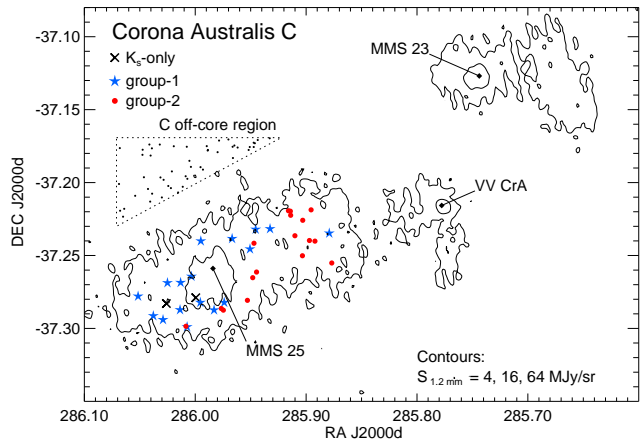}
   \caption{Sky position of the possible members
     towards MMS\,25 in field C.
     The dotted triangle encircles the stars
     in the C off-core region used for comparions.
     The off-field contains 12 group-1 and 55 group-2 stars .
   }
   \label{fig_sky_c}
\end{figure}

Furthermore, we compare the stellar surface density
per magnitude bin for on- and off-core regions.
After dereddening the stars in the colour-magnitude diagram,
the stellar surface density N
per magnitude bin is statistically indistinguishable for
on- and off-core (Table\,\ref{table_stellar_surface_density}).
Therefore, we suggest
that our near-infrared maps are able to permeate
the NW core, so that
most, if not all, of the sources detected towards
the northwestern dust ridge
are background stars.

In short, the results for region C are similar to those of NW.
Figure\,\ref{fig_sky_c} shows the sky position of the possible members
towards core C and the off-core field.
Notably, in the immediate area around the extended 1.2\,mm source
MMS\,25, 
no source is detected on our NIR maps. MMS\,25 may be slightly too
opaque for background stars to be detected on our maps, but class I-III
stars in the cloud front layer should have been seen if they exist.
Obviously, cloud core C contains at most very young proto-stars, which are
too faint and embedded to be visible at near-infrared wavelengths.

Notably, on the SIMBA map, 
the dust peaks MMS\,23 and the {\it Coronet} containing star formation
appear to have faint emission tails extending more to the west than
to the east (Figs.\ref{fig_coronet_nw} and \ref{fig_sky_c}). 
The same holds for the quiet 1.2\,mm
source MMS\,25 in cloud C. Also the C$^{18}$O maps show these tails
(Harju et al. 1993). We address possible implications in the final
conclusions (Sect. \ref{section_conclusions}).

To summarise, applying the projected  mm-excess technique to
our data yields: the {\it Coronet} contains about 53 stars ($\sim$20
known + 33 new) in an area of 40 arcmin$^{2}$ having $S_{1.2m} >
25$\,MJy/sr, yielding a stellar surface density of 1.3/arcmin$^{2}$.
In contrast,  we find no evidence of 
candidate members 
in the NW and C cores.
From the values in Table~\ref{table_stellar_surface_density}
we estimate that each of these cores contains at most 3 stars more massive
than M4, corresponding to a stellar surface density of less than
0.3/arcmin$^{2}$. This leads us to conclude that compared with the
{\it Coronet} the NW and C core regions appear to be devoid of
associated stars detectable on our near-infrared images.

\begin{table}
\begin{minipage}[t]{\columnwidth}
\caption{Comparison of
  NW and C on- and off-core fields. The double-rows list  the
  number of stars and the stellar surface density per arcmin$^{2}$ with
  Poisson errors.}
\label{table_stellar_surface_density}
\centering
\renewcommand{\footnoterule}{}  
\begin{tabular}{@{\hspace{0.1mm}}c@{\hspace{1mm}}|@{\hspace{1mm}}c@{\hspace{1mm}}|@{\hspace{1mm}}c@{\hspace{1mm}}|@{\hspace{1mm}}c@{\hspace{1mm}}|@{\hspace{1mm}}c@{\hspace{0.1mm}}}
\hline
                  & NW core& NW off& C core & C off        \\
area/arcmin$^{2}$ &  14.4            & 11.0                 & 20.9             &  13.0              \\
\hline
$   K_{s}<13$     & 12              & 13                   & 19                &  17                \\
                  & 0.83 $\pm$ 0.24 & 1.18 $\pm$ 0.33      & 0.91 $\pm$ 0.21   &  1.31 $\pm$ 0.32   \\
$13<K_{s}<15$     & 19              & 22                   & 13                &  30                \\
                  & 1.32 $\pm$ 0.30 &  2.00 $\pm$ 0.43     & 0.62 $\pm$ 0.17   &  2.31 $\pm$ 0.42   \\
$K_{s}$-only      &  5              &                      &  2                &                    \\
                  &  0.35 $\pm$ 0.16 &                      &  0.10 $\pm$ 0.07 &                    \\
\hline
\end{tabular}
\end{minipage}
\end{table}

\subsection{Search for K-band excess sources}
\label{section_disc_excess_sources}

To identify young cloud members, which are detected in all
three filters $JHK_{s}$, a well-known strategy is to look for
$K$-band excess sources by means of the colour-colour diagram
$J-H$ versus $H-K_{s}$. An excess of $K$-band emission is then
interpreted as an additional contribution from hot (T\,$\sim$\,1500\,K)
circumstellar dust, hence an indicator of pre-main-sequence objects
that are likely associated with the parent molecular cloud. This
strategy, however, requires relatively high accuracy in the
photometry, the knowledge of the NIR reddening law, and no dilution of
a $K$-band excess by scattering. So far only a 
few $K$-band excess sources have been found in the R\,Cr\,A
cloud, virtually all of them bright stars (Wilking et al.
1997) saturated on our maps.
Unfortunately, our new probable members with projected  mm-excess do not
have $JH$ photometry. Here we have examined the sources fainter than
$K_{s}$ = 11 mag. The slope of the $A_V$ direction in the
colour-colour diagrams has been derived from several sub-sets. 

\begin{figure}
   \centering
   \includegraphics[width=\columnwidth]{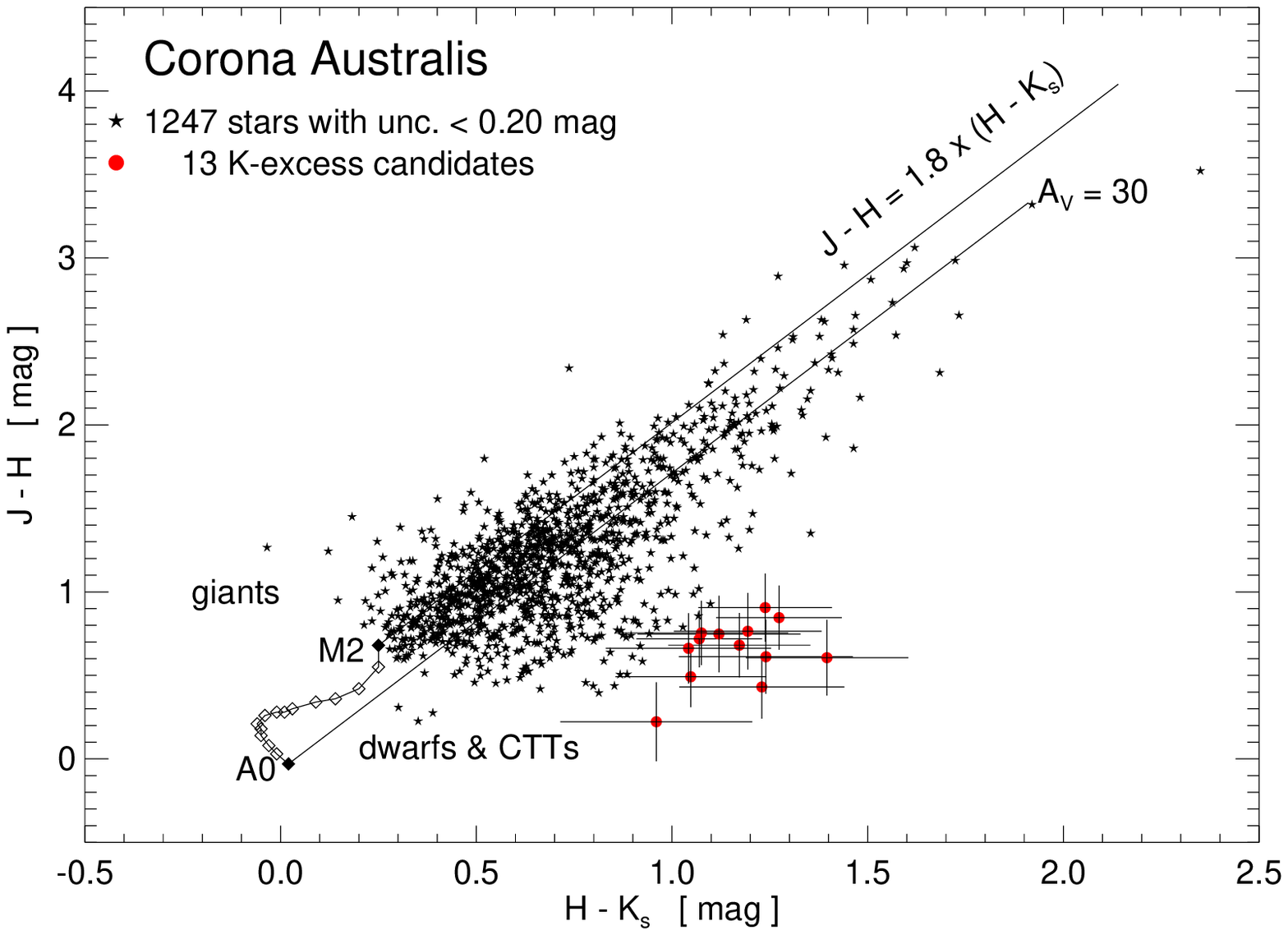}
   \includegraphics[width=\columnwidth]{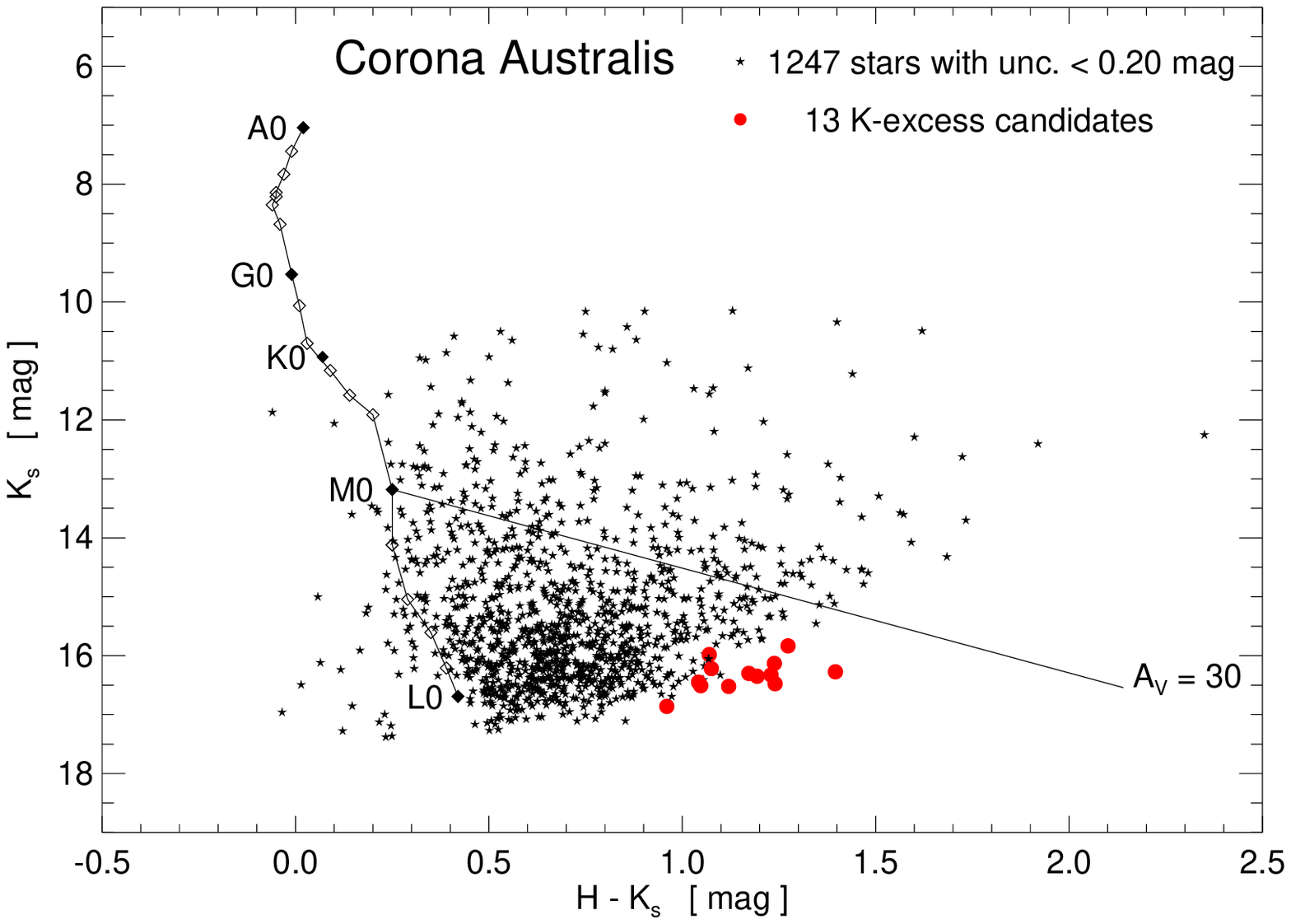}
   \caption{Colour-colour and colour-magnitude
     diagrams showing the $K$-band excess candidates. 
     The statistical significance for a $K$-band excess is low because of
     the large 1$\sigma$ error bars in the colour-colour diagram. 
     In the colour-magnitude
     diagram the $K$-band excess candidates lie at the faint end of our 
     sources. 
   }
   \label{fig_cc}
\end{figure}

To summarise, based on our $JHK_{s}$ data we find about a dozen 
sources having a small $K$-band excess compared to the reddened main
sequence (Fig.\,\ref{fig_cc}, top). However, all these sources are
faint (Fig.\,\ref{fig_cc}, bottom) and the statistical
significance for $K$-band excess among them is
marginal ($< 3\sigma$).
The photometric uncertainty as derived from the source extraction tools
does not account for additional systematic errors that may be caused by
unfavourable observing conditions. 
Therefore, we hesitate to claim the indicated $K$-band excess as real 
and call these sources $K$-band excess candidates. 
We just note that one source
(RA/DEC B1950 = 18:58:53.3 / -37:03:28) listed as a $K$-band excess
source by Wilking et al. (1997) now turns out as a double source
with 1$\farcs$6 separation at $PA = 175^{\circ}$ without evidence
of any $K_{s}$-band excess.

Our $K$-band excess candidates are located outside of the
northwestern dust ridge (3-5 sources), south of the {\it Coronet} (3-4
sources), and southeast of VV\,CrA (4 sources). If they were associated
with the R\,CrA cloud, then they would be  
very low-mass stars  (Fig.\,\ref{fig_cc}, bottom).
Notably, M-dwarfs with accretion and outflow signatures have been found 
3$\farcm$5 and 5$\farcm$3 south and southwest 
of the {\it Coronet} by means of slitless optical 
spectroscopy and detailed follow-up studies 
(Fern\'andez \& Comer\'on 2001, 2005; Barrado y Navascu\'es et al. 2004). 
These very low-mass stars, lying outside the region of bright 
1.2\,mm emission, do not show any signs of K-band excess, either. 

Obviously our NIR sources
do not exhibit circumstellar discs with prominent hot dust
emission showing up already in the $K$-band. 
Sources with cooler circumstellar discs 
may be found by combining our data with longer wavelength
Spitzer data.

\subsection{Stars with nebulae and Herbig-Haro objects}
\label{section_nebulae}

\begin{figure}
   \centering
   \includegraphics[width=\columnwidth]{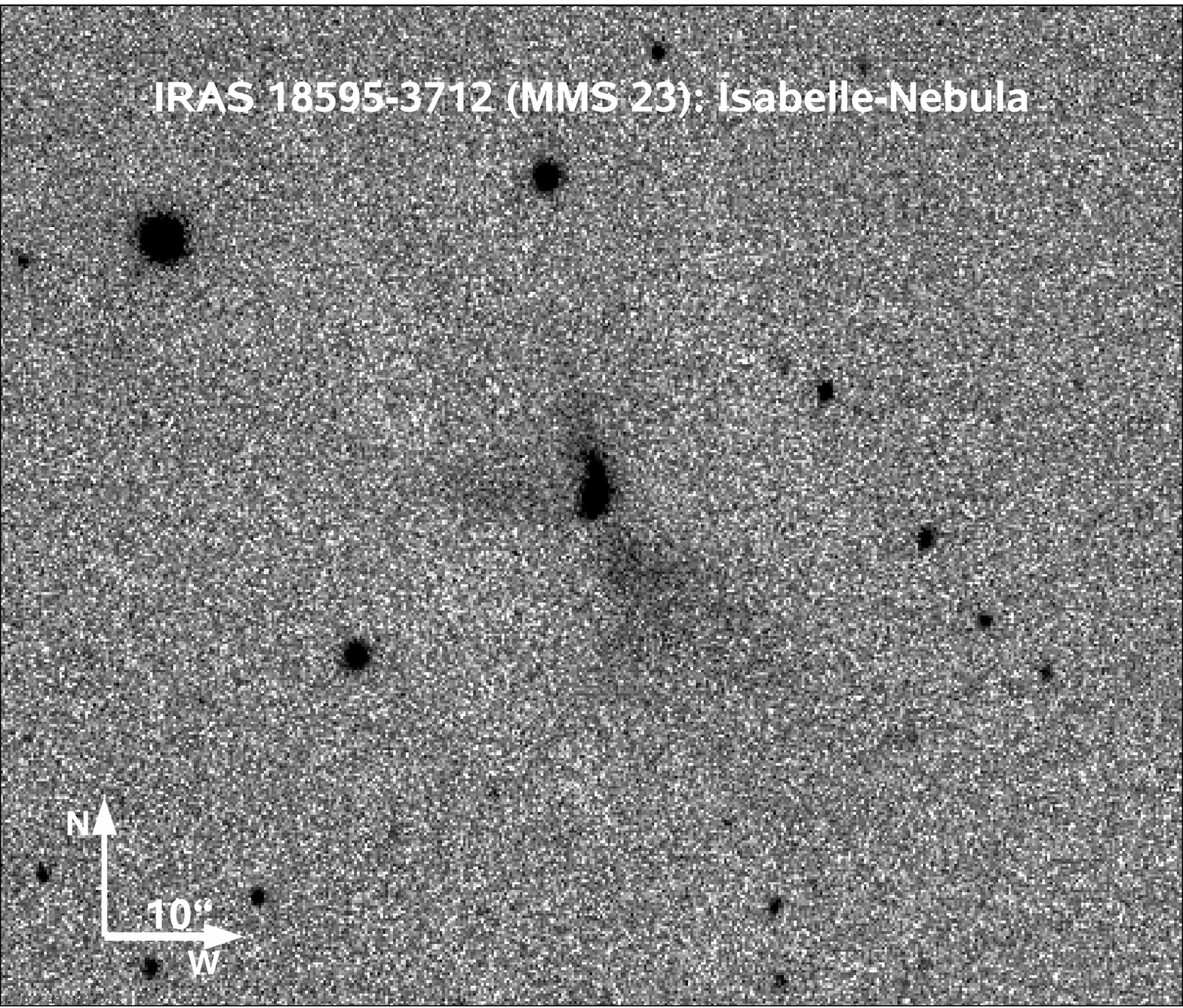}
   \includegraphics[width=\columnwidth]{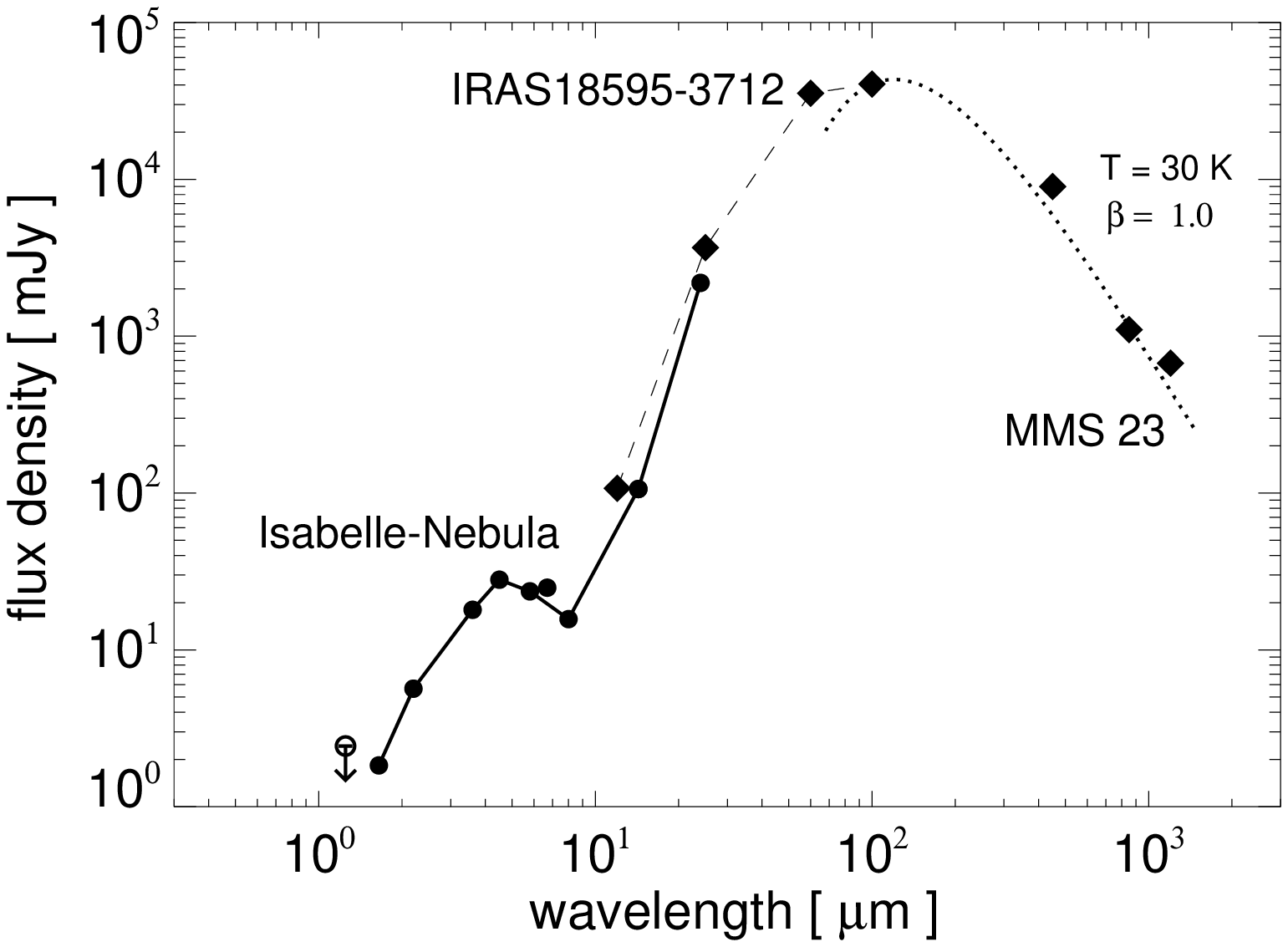}
   \caption{$K_{s}$-band image and SED of the X-shaped
     Isabelle Nebula.
     The lines and symbols in the SED plot mean: filled circles
     connected by the solid line mark our NIR data together with
     Spitzer and ISOCAM MIR data; the filled diamonds connected with a
     long-dashed line mark IRAS data; SCUBA 450 +  850 $\mu$m and
     SIMBA 1.2 mm photometry is plotted with filled diamonds; the dotted line
     represents a modified blackbody with emissivity index 
     $\beta = 1$ and temperature $T = 30 K$.  
   }
   \label{fig_isabelle_nebula}
\end{figure}

Apart from planetary nebulae, nebulous and extended objects are not
expected among field stars. Hence such objects are probably
associated with the star-forming molecular cloud.

\begin{figure}
   \centering
   \includegraphics[width=\columnwidth]{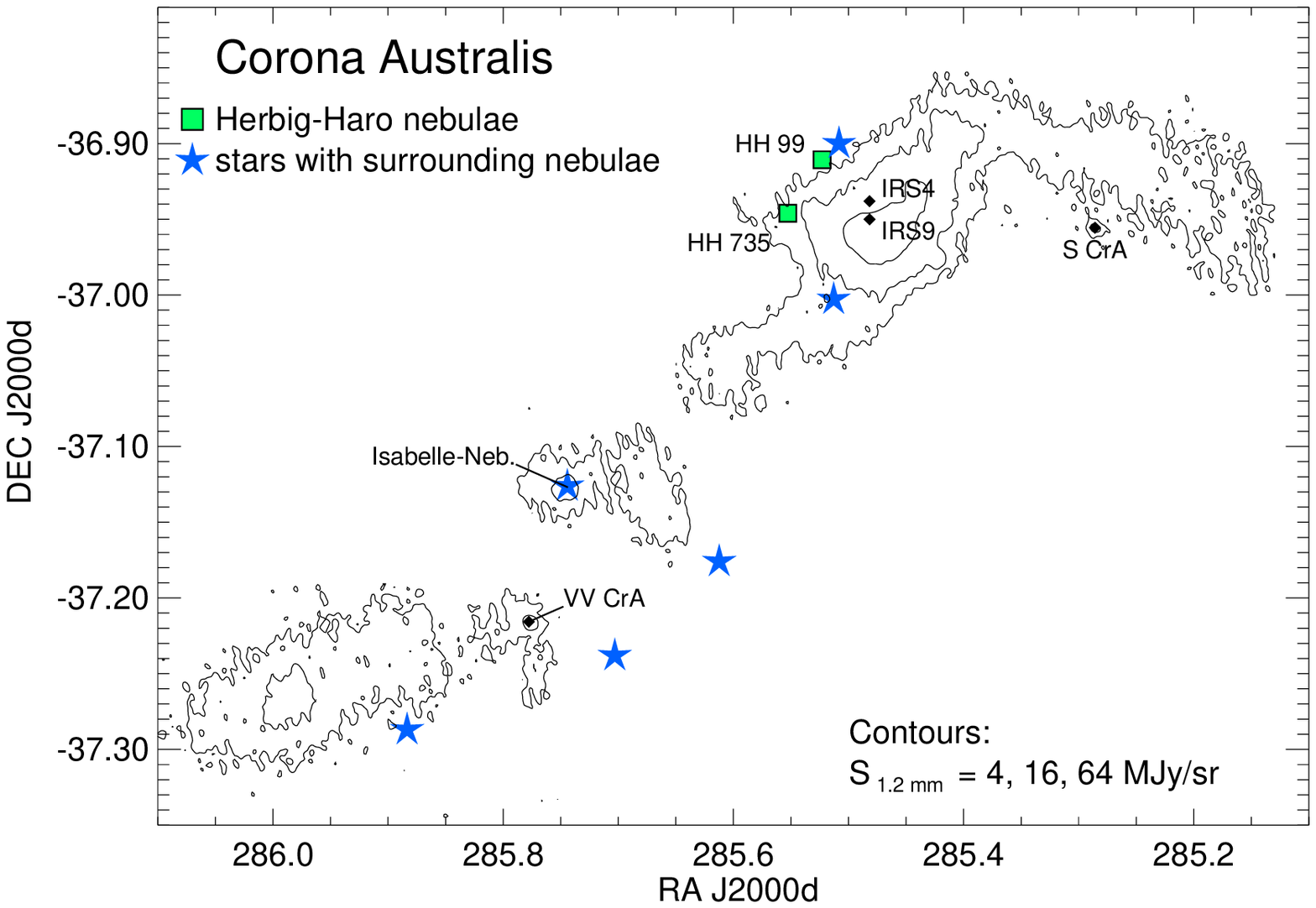}
   \caption{Sky positions of the resolved nebulae.
   }
   \label{fig_nebulae}
\end{figure}

The most conspicuous extended object is a beautiful X-shaped nebula,
hereafter referred to as the ``Isabelle-Nebula''. With an extent of
about 20$\arcsec$ it is visible in $H$ and $K_{s}$ but not in $J$.
Figure~\ref{fig_isabelle_nebula} (top) shows the $K_{s}$-band image.
The sky-projected axis of the bipolar nebula is about perpendicular
to the long axis of the entire R\,CrA cloud. 
While this could be a chance coincidence, 
a relation with the cloud's overall angular momentum also appears as an
attractive explanation.  
The north-eastern part of
the nebula appears somewhat brighter than the south-western part and
has a prominent elongated emission along its northern edge. This
bright spot has already been detected and denoted as IRAS 32c by Wilking et
al. (1992). The position listed in Table \ref{table_nebulae} refers to
the centre of the X-shape. It is coincident with
the prominent millimetre point source MMS\,23 (Chini et al. 2003)
alias IRAS18595-3712 located about 16$\farcm$3 southeast of
R\,Cr\,A. Figure~\ref{fig_isabelle_nebula} (bottom) shows the
$1-1200\,\mu$m spectral energy distribution (SED). The SED is
complemented in the
mid-infrared by ISO 6.7 and 14.3\,$\mu$m photometry (Olofsson et al.
1999) and photometry at $3.6 - 24\,\mu$m was derived from the Spitzer
archive 
using the basic calibrated data (BCD) products version S16. 
In addition, far-infrared IRAS data are shown, supplemented
by data points from SCUBA at 450 and 850\,$\mu$m (Nutter et al.
2005) and SIMBA at 1200\,$\mu$m (Chini et al. 2003). At first
glance, the integrated
$1 - 1200\,\mu$m luminosity of about 2.1\,$L_{\odot}$ suggests a
solar mass PMS object. Obviously, the star exciting the
Isabelle Nebula is deeply hidden and not directly visible at
wavelengths shorter than 10\,$\mu$m.
The SED longwards of 10\,$\mu$m may mimic a pure class I object.
However, the regular X-shape of the
Isabelle Nebula suggests obscuration by a disc-like dust
distribution seen roughly edge-on. The near-mid-infrared ($2 -
8\,\mu$m) hook of the SED cannot be interpreted as direct stellar
radiation; it is instead produced by scattered light from the
nebula.
Furthermore, if the $H$- and $K_{s}$-band fluxes are dominated by continuum
and not by line emission, the SED of the nebula appears to be
strongly reddened by foreground dust. 
When disentangling such a non-spherical geometry,
the actual luminosity of the PMS object increases significantly,
because 1) the dust covering factor of the disc as seen from the
central star is smaller ($\sim$\,50\%) than that of the sphere,
2) the scattered near-mid-IR light comes from a narrow bi-cone
with opening half-angle 30$^\circ$ illuminated by the star, and
3) the near-mid-IR nebula is strongly reddened.
Altogether, the actual luminosity increases potentially by a factor
4-10, resulting in about 8-20\,$L_{\odot}$, consistent with 
a transition Class I-II object of spectral type F
seen edge-on. 
While our coarse luminosity estimate is in line with
results from SED fitting tools 
(5-10\,$L_{\odot}$ at intermediate inclinations, Robitaille et al. 2007), 
future NIR spectra and  radiative transfer models
constrained to the edge-on case 
will be needed to make any strong statements about the nature 
of this PMS star. 

\begin{figure}
   \centering
   \includegraphics[width=\columnwidth]{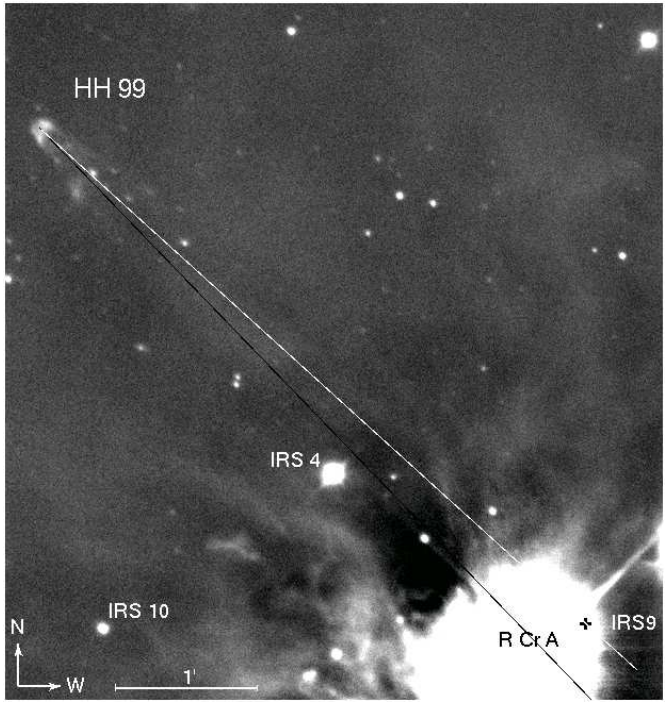}
   \caption{$K_{s}$-band image of HH\,99 with its bow-shock arrow cap.
     The black and white lines point to its driving star candidates
     R\,CrA and IRS\,9 (marked with a "$+$" sign, right of R\,CrA), 
     respectively. 
     Because the bright stars are saturated in our frames, the position of the
     sources has been taken from Wilking et al. (1997) and Taylor \&
     Storey (1984). The positional uncertainty is about the size of the
     "$+$" sign.
   }
   \label{fig_hh99}
\end{figure}

Another five objects fainter than $K_{s} = 10\,$mag show small
(3-5$\arcsec$) elongated halos around the central star.
They could not be resolved further in our images.
At the distance of R\,CrA their NIR brightness indicates
subsolar luminosities. We confirm the two sources listed as
nebulous by Wilking et al. (1997). The sky position
of the nebulae is shown in Fig.~\ref{fig_nebulae}.
In addition to the faint diffuse nebulae, we note that
the bright Herbig Ae/Be-star TY\,CrA exhibits a remarkably sharp
sickle-like double shell indicating multiple outbursts. 

When looking at Herbig-Haro objects, HH\,99 has a clear bow-shock
arrow cap (Fig.~\ref{fig_hh99}). Drawing a mid-axis through the
regular shape of HH\,99 suggests that either R\,Cr\,A or IRS\,9 is its
energy 
source. Because R\,Cr\,A can be excluded as discussed by Wilking et
al. (1997), we suggest that IRS\,9 is the most likely driving source
for HH\,99. IRS\,6 (located about 50$\arcsec$ 
 northwest of R\,CrA)
as the driver of HH\,99, as proposed by Wang et al. (2004), may be
questioned on the basis of alignment. 

In general, the $K_{s}$-band flux of the Herbig-Haro objects is
probably dominated by the shocked molecular hydrogen line
H$_2$\,1-0\,S(1) at 2.121\,$\mu$m and not by hot dust emission
(e.g. Wilking et al. 1997, McCoey et al. 2004). Similarly, the $J$ and
$H$-band covers the FeII lines at 1.257 and 1.644\,$\mu$m,
respectively. Our deep $K_{s}$-band image of field A shows several
more Herbig-Haro objects. For example, HH\,732\,A-C, located about
5$\arcmin$ north of R\,Cr\,A and HH\,735, 
located about 25$\arcsec$ north of the SIMBA point source MMS\,19.
Based on
alignment with the [SII]\,$\lambda$6763\AA\ contours elongated at PA
220$^{\circ}$, Wang et al. (2004) suggested that HH\,735 is driven by
IRS\,7 located 3$\farcm$5 at $PA = 260^{\circ}$. However, the
optical data may suffer from patchy extinction. Drawing a mid-axis
to the quite symmetric sickle shape seen in all three near-infrared
bands points towards IRS\,4 as a potential driving source.
Most detected Herbig-Haro objects are located outside the area of
bright 1.2\,mm emission ($S_{1.2}> 25~MJy/sr$),
hence in a medium of low-moderate dust density.
HH-861, HH-861, and HH863 detected with some
caveats by Wang et al. (2004) around VV\,CrA are outside the FOV of
our NIR maps in the southeastern part.

To summarise, while Herbig/Haro objects appear to be mainly
concentrated around the coronet cluster, some faint stars with
(compact) nebulae are found also towards the southeastern part of the 
elongated Corona Australis cloud (Fig.~\ref{fig_nebulae}).

\section{Conclusions}
\label{section_conclusions}

We performed a deep near-infrared survey of the entire R\,CrA
molecular cloud and analysed the data with respect to the 1.2\,mm dust
continuum map. 

\begin{itemize}
\item[1)]
Applying the projected mm-excess technique we find 33 new sources
deeply embedded in the front layer towards the {\it Coronet}. 
Nine sources could be detected at $H$ and $K_{s}$, 24 only have
$K_{s}$-band data. In contrast to the {\it Coronet}, the NW dust ridge
and the C core region appear to be devoid of associated stars
detectable in our near-infrared images. This strengthens the evidence
for the {\it Coronet} being the region where most of the Class I-III
stars are found. Obviously, core C contains at most very young
protostars, which are too faint and embedded to be visible at
near-infrared wavelengths. 
Stringent estimates about the contamination from background objects may be
derived from Spitzer MIR data. 

\item[2)]
Only marginal evidence of $K$-excess sources can be
found from our data ($K_{s} > 10\,$mag). 
If these $K$-excess candidates are associated with the R\,CrA
cloud, then they are very low-mass stars.
In general, it seems that our sources do not
exhibit circumstellar discs with {\it prominent} hot dust emission showing
up already in the $K$-band. 

\item[3)]
The four new nebulae found towards the southeastern part corroborate
the picture that star formation has also started in these regions.

\item[4)]
The dust and CO peaks A {\it Coronet} and B (MMS\,23) with ongoing
star formation exhibit relatively sharp eastern borders and shallow
western tails; the same holds for cloud C. 
If the star formation in the diverse R\,CrA subcondensations were
directly triggered by the wind from the Upper Centaurus-Lupus OB
association, such a morphology is not expected at first glance.
Instead in a scenario where an external wind hits an originally
symmetric gas clump, one would expect that the front side, where the
wind hits the clump, appears compressed and houses the first and stronger
star formation. Because this is not observed, the picture is
supported that the expanding Upper Centaurus-Lupus HI shell just may 
have led to fragmentation of the R\,CrA cloud into subcondensations,
which further collapse to form stars mainly under the control of their own
gravitation. 

\end{itemize}

Our results are in line with the predictions by Harju et al.
(1993) that the R\,CrA cloud has fragmented into subcondensations
that are at different star-forming stages.

\begin{acknowledgements}
We thank Katrin K\"ampgen for preparing the observations, Nicole van
der Bliek and Marcus Albrecht for performing the observations,
and Markus Nielbock for intriguing discussions, in particular
on the Herbig-Haro objects and nebulae. We also thank Bruce Wilking
and Jochen Eisl\"offel for sending us their electronic $JHK$ and
$RI$-band data files for comparison. The analysis greatly benefitted
from the Aladin tool. 
This publication makes use of data products from the Two Micron All
Sky Survey, which is a joint project of the University of
Massachusetts and the Infrared Processing and Analysis
Center/California Institute of Technology, funded by the National
Aeronautics and Space Administration and the National Science
Foundation.
We thank the anonymous referee for the constructive, detailed report. 
This work was supported by the
Nordrhein--Westf\"alische Akademie der Wissenschaften. 
\end{acknowledgements}

\end{document}